\documentclass[%
 reprint,
 amsmath,amssymb,
 aps,
]{revtex4-2}

\usepackage{graphicx}
\usepackage{dcolumn}
\usepackage{bm}
\usepackage{multirow}
\usepackage{color}
\usepackage[normalem]{ulem}
\usepackage{appendix}
\definecolor{colortodo}{RGB}{255,0,0}

\begin{document}

\preprint{APS/123-QED}
\title{ 
Escape dynamics of confined undulating worms
}

\author{Animesh Biswas}
\author{Arshad Kudrolli}%
 \email{Corresponding author email: akudrolli@clarku.edu}
\affiliation{Department of Physics, Clark University, Worcester, MA 01610}%

\date{\today}

\begin{abstract}
We investigate the escape dynamics of oligochaeta {\it Lumbriculus variegatus}  by confining them to a quasi-2D circular chamber with a narrow exit passage. The worms move by performing undulatory and peristaltic strokes and use their head to actively probe their surroundings. We show that the worms follow the chamber boundary with occasional reversals in direction and with velocities determined by the orientation angle of the body with respect to the boundary.  The average time needed to reach the passage decreases with its width before approaching a constant, consistent with a boundary-following search strategy. We model the search dynamics as a persistent random walk along the boundary and demonstrate that the head increasingly skips over the passage entrance for smaller passage widths due to body undulations. The simulations capture the observed exponential time-distributions taken to reach the exit and their mean as a function of width when starting from random locations. Even after the head penetrates the passage entrance, we find that the worm does not always escape because the head withdraws rhythmically back into the chamber over distances set by the dual stroke amplitudes. Our study highlights the importance of boundary following and body strokes in determining how active matter escapes from enclosed spaces.      
\end{abstract}

\maketitle

\section{Introduction}
Motile organisms navigate complex natural habitats with heterogeneous structure and physical boundaries in search of food and shelter~\cite{rizkalla2007explaining,Jung2010,bilbao2013nematode,hosoi15,kudrolli2019burrowing,bhattacharjee2019bacterial}. While light and sound are widely used to navigate complex environments by higher organisms~\cite{alexander2003principles}, mechanosensation can be the primary sense used to navigate dark subterranean environments. Organisms overcome obstacles that are too large to push aside by finding openings that allow passage. How organisms achieve this from local knowledge of topography can shed light not only on strategies learned by the organisms through evolution, but allow one to identify the general physical principles in play~\cite{bechinger2016active}. This understanding can be useful in designing autonomous systems optimized to operate in such environments, or even assess the physical properties of the medium~\cite{volpe2011microswimmers,biswas2020first,trivedi2008soft,winter14}.
\vspace{0.05in}

The diffusion of bacteria and other self-propelled organisms through disordered porous media have been studied with experiments and active filament models~\cite{kurzthaler2021geometric,irani2022dynamics,bhattacharjee2021chemotactic,ford2007role}. 
It is well established that microswimmers can align and move along boundaries due to hydrodyanmic interactions leading them to aggregate, rather than moving around uniformly, depending on the topology of the environment~\cite{Park2003,park2008enhanced,lauga2016bacterial,ohmura2018simple,denissenko2012human,berke2008hydrodynamic,yuan2015hydrodynamic,berke2008hydrodynamic,chopra2022geometric}. However, interactions mediated by the interstitial medium can be less important in larger organisms, and direct contact by touching needs to occur to identify features. At the most rudimentary level, such an interaction with a solid boundary can be considered as steric. Depending on its rotational diffusion and shape, the organism can still remain trapped at the boundary or reflect freely from the surface~\cite{elgeti2016microswimmers}. Even in microorganisms, cilia and flagella are known to directly detect surfaces in swimming eukaryotes~\cite{kantsler2013ciliary,belas2014biofilms,thery2021rebound,sipos2015hydrodynamic}. And, mechanosensory neurons are known to guide nematode~\textit{C. elegans} as they move in soil~\cite{li2011neural}. Many large organisms, such as cockroaches~\cite{jeanson2003model,creed1990interpreting}, ants~\cite{dussutour2005amplification}, fish~\cite{schnorr2012measuring}, rodents~\cite{treit1988thigmotaxis}, and humans~\cite{walz2016human} have been noted to exhibit boundary following behavior when they come into contact with boundary walls. This thigmotactic behavior involves moving along the edges of surfaces, such as walls or other boundaries, often in a repetitive pattern influenced by sensory feedback, and environmental cues. Dorgan, and collaborators have investigated the burrowing dynamics of various annelid and opheliid organisms in muddy and sandy environments~\cite{dorgan2018kinematics,dorgan2005burrow, francoeur14}. They demonstrated how these organisms interact with their surroundings to create and maintain their burrows. Nonetheless, how exactly steric interactions can be employed to find passages and efficiently navigate randomly structured medium is not widely explored, not least due to the difficulty in observing organisms in such environments.  Further questions arise whether the body strokes used in achieving motion and collisional contact with the boundary can interfere with identifying boundary features. The study of motile organisms where the entire body and its shape are tracked are few and can lead to deeper understanding of strategies used to navigate through tight spaces. 

The freshwater oligochaete {\it Lumbriculus variegatus} is commonly found across temperate regions in North America and Europe in sediment beds at the bottom of water bodies, and is increasingly employed as a laboratory model due to its well-established biological facts, macroscopic size, and easy maintenance~\cite{brinkhurst1991annelida,drewes1999helical,Lesiuk2001,sciadv.abj7918,phipps1993use,martinez2021cuts}. Their long slender limbless body is representative of many organisms which move underground, and has also been used to study  collective dynamics and physicochemical behavior~\cite{ozkan2021collective,tuazon2022oxygenation,nguyen2021emergent}.  These  worms have sensory nerves all over their bodies that help them respond to threats and detect obstructions in their path. Unless provoked, they typically move in the direction of their head, employing transverse undulatory and peristaltic strokes depending on the rheology of the medium~\cite{kudrolli2019burrowing}. 

\begin{figure*}
\centering
\includegraphics[width=0.6\textwidth]{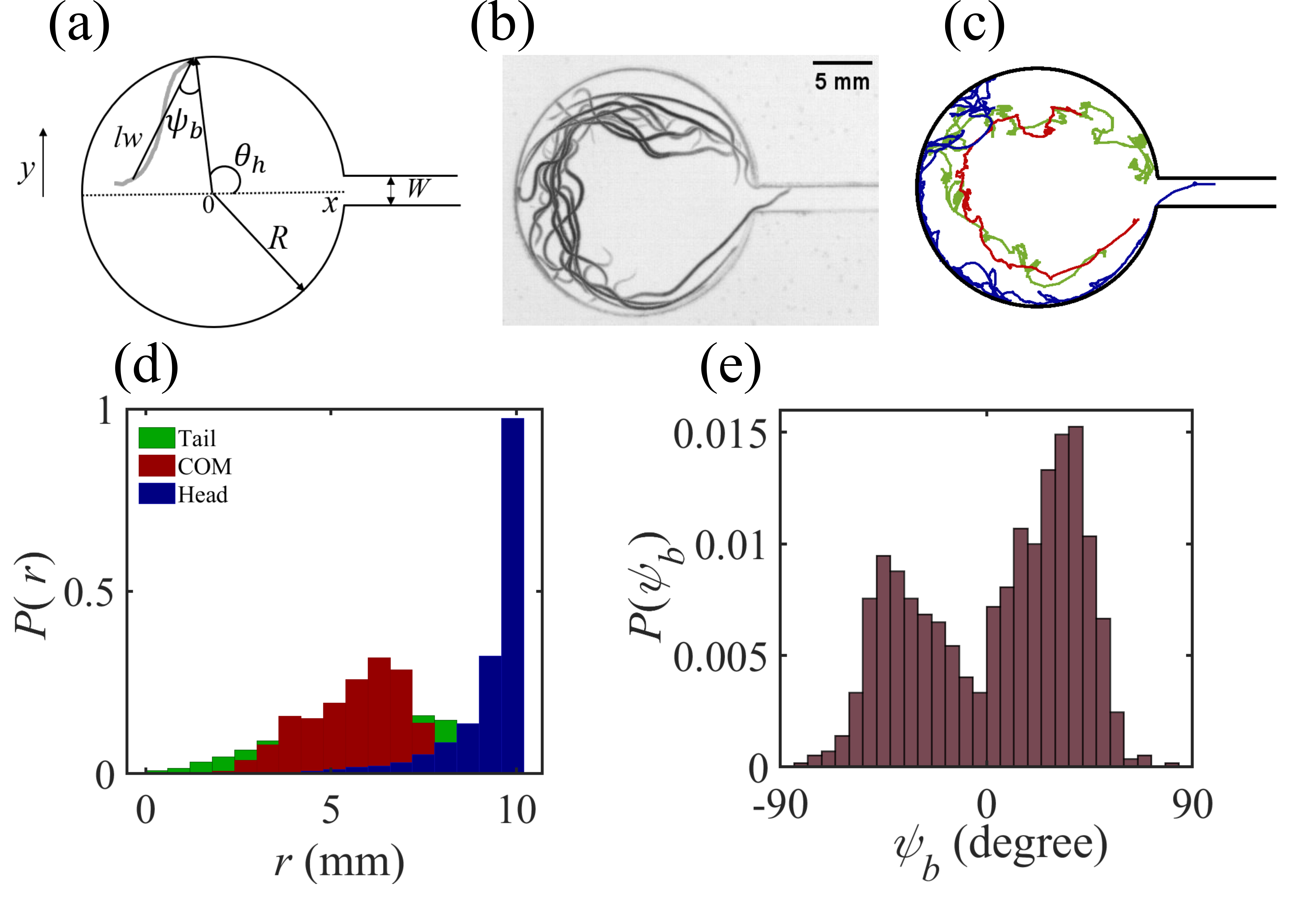}
\caption{\label{fig:chamber} (a) A schematic of the circular enclosure with an exit passage with radius $R$ and exit width \textit{W}. $\theta_h$ is the angular position of the head from the center O, $\psi_b$ is the body orientation angle with the normal, and $l_w$ is the length of the worm. (b) The motion of the worm in the corresponding experimental system is illustrated by superposed images taken at 1 frame per second. The worm can be observed to stay close to the boundary. (c) The tracked trajectories of head (blue), centroid (red), and tail (green) of the worm as it follows the boundary. (d) The probability distribution $P(r)$ of radial positions $r$ of the worm's head ($r_h$), centroid ($r_c$), and tail ($r_t$) in the chamber from the center ($R = 10$\,mm; $W = 2$\,mm). (e) The distribution of body orientation angle $\psi_{b}$. The data averaged over 20 trials are plotted in (d) and (e).}
\end{figure*}

Here, we examine the dynamics of {\it L. variegatus} as they move inside a transparent water-filled chamber connected to an exit passage which is much smaller than their lengths, but also much wider than their widths. We visualize and track their dynamics as they move around the chamber and escape through the exit passage repeatedly. Their motion is studied by measuring the body orientation and detailed interaction of the various parts of their body with the boundary and passage. We find that the worm does not explore the chamber uniformly, but rather follows the chamber boundary guided by the interaction of its head with the boundary and its entire body orientation. Then, we focus on the time scales needed to find the passage and escape as a function of the exit width, and illustrate the importance of the body fluctuations on success rate. A minimal boundary search random walk model is developed to capture aspects of the observed behavior.

\section{Methods}
Figure~\ref{fig:chamber}(a) schematically shows the system consisting of a horizontal circular chamber 
with a narrow exit passage in which we study the escape dynamics.  In practice this is accomplished by laser cutting out two circular chambers of equal radius $R$  from an Acrylic sheet of thickness $h =1.5$\,mm and connecting them with a narrow straight passage of length $L$ and width $W$. This sheet is then sandwiched between removable top and bottom transparent sheets to confine the worm to quasi-2D while fully immersed in a water bath. 

We perform experiments with {\it L. variegatus} of length $l_w$=20 $\pm 5$\,mm, diameter $d_w \approx 100\,\mu$m and systems with $R =10$\,mm and $L=35$\,mm to reduce the number of experimental parameters. These worms move typically with speeds $v_w \approx 2$\,mm/s, with longitudinal peristaltic strokes with amplitude $A_L \approx 1$\,mm and time period $T_L \approx 0.6$\,s, and transverse undulatory strokes with amplitude $A_T \approx 1.4$\,mm and time period $T_T \approx 12$\,s~\cite{kudrolli2019burrowing}. Further details on maintaining them are similar to those in Ref.~\cite{kudrolli2019burrowing}. The effect of system geometry on escape rate is investigated by using $W = 0.25$\,mm, 0.5\,mm, 1\,mm, 2\,mm, 3\,mm, 4\,mm, and 5\,mm. Thus, $W$ varies from being less than, to greater the stroke amplitudes, while being always much greater than the worm diameter, and much less than $L$. Since the chamber floor is horizontal, this range of $W$ leads to an entropic barrier for escape, rather than a physical one. \textit{L. variegatus} were  obtained  from  Carolina  Biological  Supply Company  (https://www.carolina.com)  on  October  3,  2017. The worms are sustained  according to suppliers' specifications in a freshwater aerated aquarium in a laboratory  under ambient lighting with a HVAC system which maintains the temperature at $24\pm2^\circ$C. The transferring of the worm into the observation chamber was performed with a plastic pipette which disturbed the worms minimally.

The worm and the enclosure are imaged with a megapixel digital camera from above with back illumination which causes the worm and system boundaries to appear dark against a bright background. The overall intensity of these lights is not different from the ambient lab lighting. The entire body including the head and tail is tracked by image processing~\cite{kudrolli2019burrowing} over the 30\,minutes time duration of a typical trial. Figure~\ref{fig:chamber}(a) further shows the coordinates used to denote the worm's position and orientation w.r.t. the $x$-axis located at the chamber center. The axis origin is located at cell center and directed toward the center of the exit passage, the distance of the worm's head from the chamber center $r_h$, its angular position $\theta_h$ from the $x$-axis, and body orientation angle $\psi_b$ relative to the normal to the surface are also denoted in Fig.~\ref{fig:chamber}(a).

\section{Results}

\subsection{Motion in the chamber}
When a worm is placed inside a chamber, it begins to move after a typical acclimatization time of a few seconds, and explores the system by interacting with the boundary, locating the passage between the two chambers, and passing back and forth several times between the chambers (see Movie~S1).  Thus, focusing on the worm after its initial acclimatization time, we have a system as represented by Fig.~\ref{fig:chamber}(a), where the worm moves around a chamber till it locates the passage, and exits into the other chamber, where the process repeats itself.  We observe the motion of the worm over at least 30~minutes, and perform at least 50 trials for each $W$ to obtain statistically significant information on the time taken by the worm to find and enter the passage. The water was typically replaced after 2 to 3 trails to maintain the water quality.

Figure~\ref{fig:chamber}(b) shows superimposed images of the worm as it moves around the chamber and exits through the passage. We plot the corresponding tracked path of the worm's head, tail, and centroid in Fig.~\ref{fig:chamber}(c), and observe that the head appears to be in almost constant contact with the boundary. The tail and the centroid in contrast stay away from the boundary. Fig.~\ref{fig:chamber}(d) shows a plot of the measured probability distribution of the radial position of the head $r_h$, tail $r_t$, and centroid $r_c$ of the worm while moving in the chamber, averaging over data from 20 trials over approximately ${{1107}}$\,seconds. Even over this larger sample set, we observe that the head spends a significant fraction of time near the boundary, while the tail wanders more broadly within the chamber. Further, one observes that the centroid is typically located at $r_c/R \approx 0.6$, showing that the worm is not uniformly distributed inside the chamber even over extended periods, and $r_h/R \approx 1$ shows the worm spends most of the time exploring the boundary.

Plotting the probability distribution of the angle  $\psi_b$ that the body makes with the normal to the boundary at the head location in Fig.~\ref{fig:chamber}(e), we find that it is not uniform, but rather has two peaks, broadly distributed with means at approximately $\psi_b = -34$\,degrees and $30$\,degrees. Considering the width of their distribution, these angles as well as the peak of $r_c$ in $P(r)$, denote that the worm typically moves clockwise or counter-clockwise around the boundary circular chamber, while its centroid is located at a constant distance from the boundary. Thus, while {\it L. variegatus} have sensory nerves all over their body, and react to touch by recoiling and moving away rapidly~\cite{drewes1999helical}, they appear to primarily use their prostomial nerves to probe physical obstacles and move forward. The rest of the body appears to react essentially passively to the forces experienced when it touches boundaries.

\subsection{Boundary interactions}
\begin{figure}
\centering
{\includegraphics[width=0.5\textwidth]{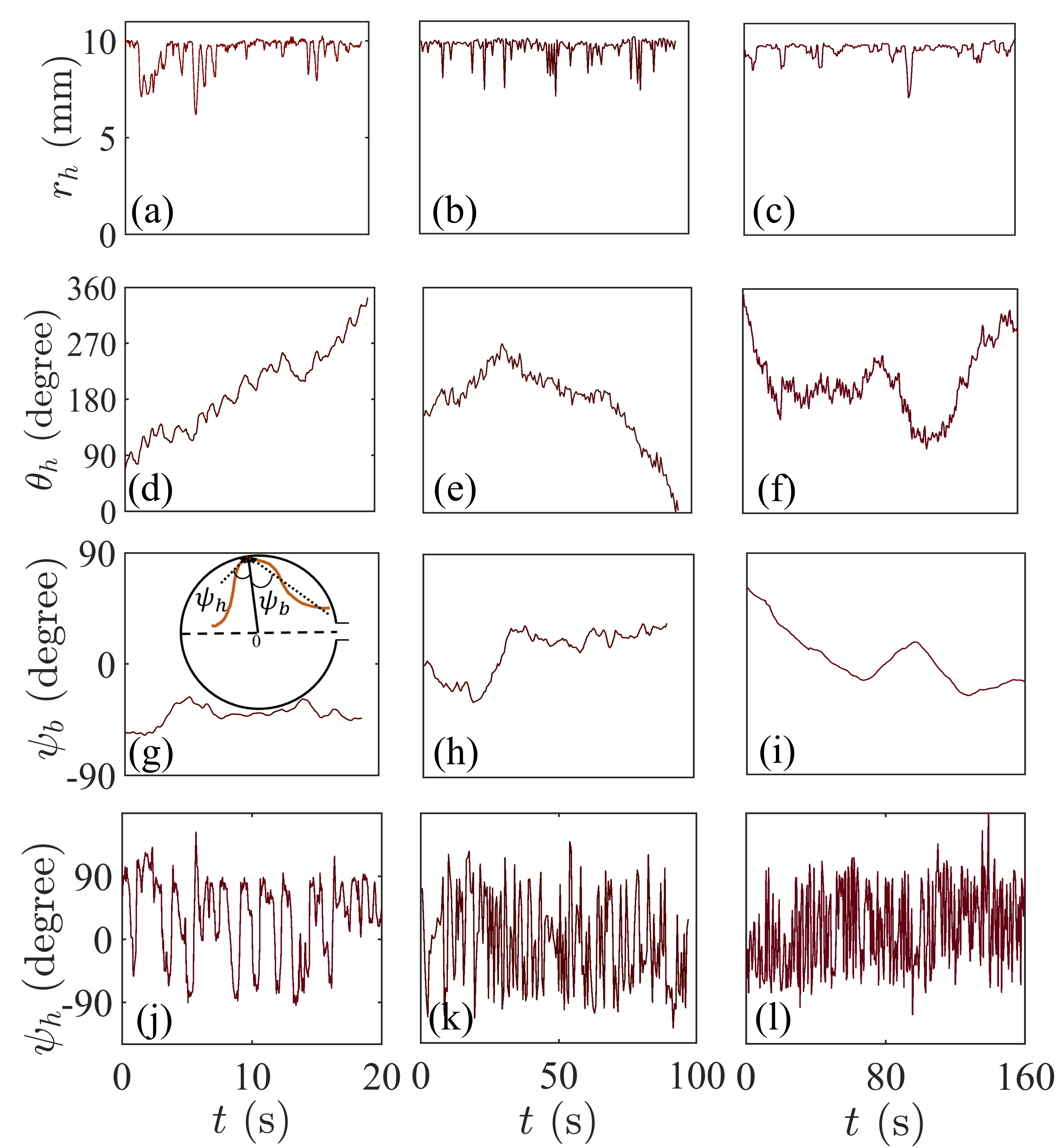}}
\caption{\label{fig:theta_psi} Examples of the radial position of the head $r_h$ (a-c), angular position of the head $\theta_h$ (d-f), body orientation $\psi_b$ (g-i) and head orientation $\psi_h$ (j-l) as a function of time $t$. The worm moves clockwise (a), clockwise and then switches to counterclockwise (b), and as it changes direction several times (c). The direction of motion changes when $\psi_b$ changes sign, but appears uncorrelated with $\psi_h$, which fluctuates more widely.Inset to (g): Schematic shows the body orientation angle $\psi_b$ and head orientation angle $\psi_h$}.
\end{figure}

To gain a deeper understanding of the interaction of the worm with the boundary, we plot three examples of the radial position of the head from the chamber center $r_h$  in Fig.~\ref{fig:theta_psi}(a-c), its corresponding angular position $\theta_h$ in Fig.~\ref{fig:theta_psi}(d-f), body orientation angle $\psi_b$ in Fig.~\ref{fig:theta_psi}(g-i). We also plot the head orientation angle $\psi_h$ that the head subtends with the normal to the boundary in Fig.~\ref{fig:theta_psi}(j-l). In all three examples, the position and the orientation of the worm is plotted from when it first touches the boundary till it approaches/enters the exit passage. In each example, $r_h$ remains close to $R$, except for very short periods, when it is not in  contact as the head moves rapidly away and then towards the boundary. 

\begin{figure*}
\centering
\hspace*{-0.7cm}    
{\includegraphics[width=0.65\textwidth]{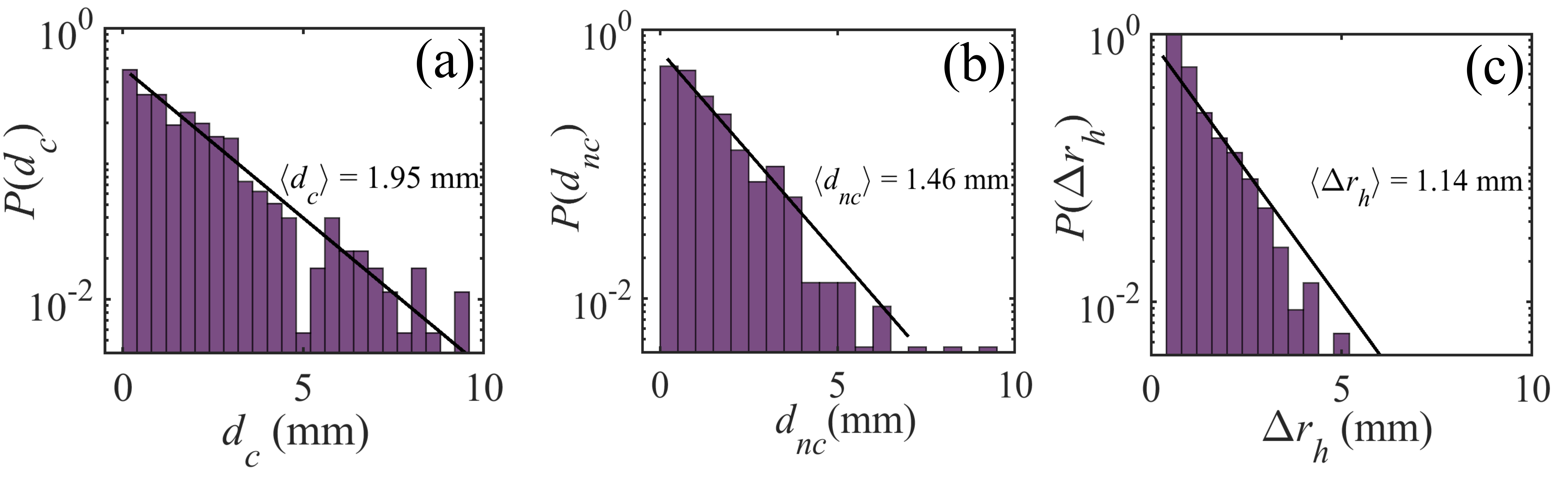}}
\caption{\label{fig:scale} (a) The distribution of distance $d_{c}$ over which the worm travels while it is in contact with the boundary. (b) The distribution of distance $d_{nc}$ over which the worm travels when it loses contact with the boundary. (c) The distributions of radial distance traveled by the head $\Delta r_h$ when not in contact with the boundary. The distributions can be described by exponential functions with means of order of the dual stroke amplitudes $A_L$ and $A_T$.}
\end{figure*}

\begin{figure*}
\centering
{\includegraphics[width=0.7\textwidth]{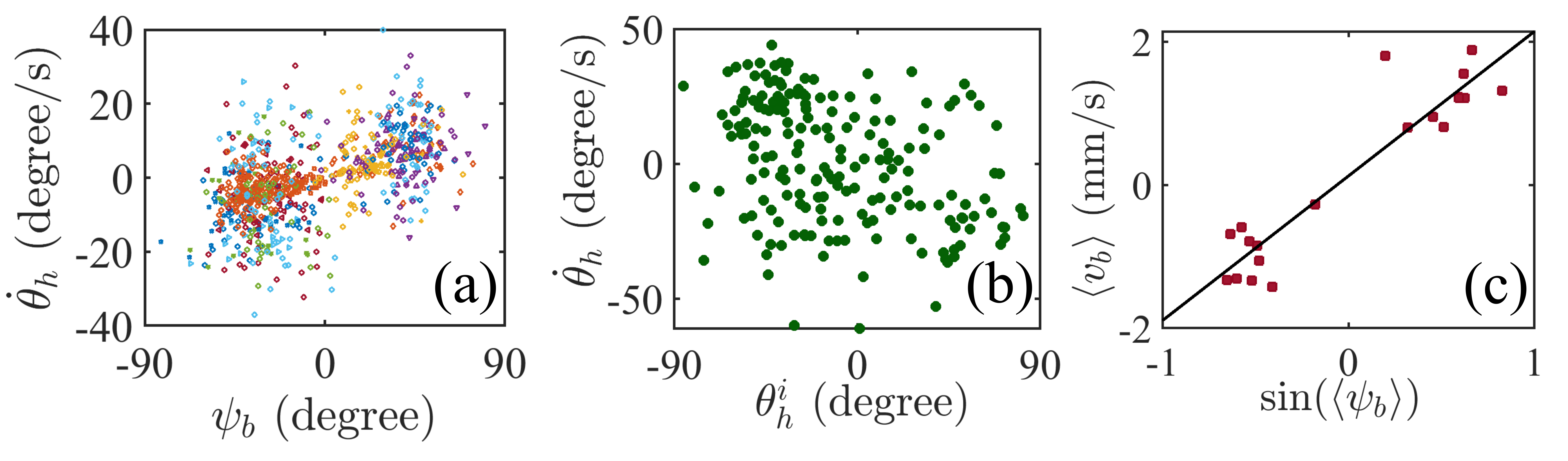}}
\caption{\label{fig:interaction}  (a) The angular speed of the worm is correlated with incident body orientation angle. (b) Angular speed of the worm  $\dot{\theta_h}$ after a collision as a function of head incident angle. No correlations are evident. (c) The mean tangential boundary velocity $\langle v_b \rangle$ as a function of $\sin({\langle \psi_b \rangle})$. A line fit given by $\langle v_b \rangle = v_w \sin({\langle \psi_b \rangle})$ is shown. The fitted value $v_w=2.02$\,mm/s, corresponds to the unobstructed speed of the worm.}  
\end{figure*}

Figure~\ref{fig:scale}(a, b) shows the distribution of distances along the boundary traveled by the worm's head when it is in contact with the boundary $d_{c}$ (using a criteria $R - r_h < 0.3d_w$ to define contact,) and when it travels while not in contact with the boundary $d_{nc}$.   We observe that the distributions corresponding to these distances travelled along the surface while in contact, and while not in contact, can be described by exponential functions, $P(d_{c}) = \frac{1}{\langle d_c \rangle} \exp{(-d_{c}/\langle d_c \rangle)}$ and $P(d_{nc}) = \frac{1}{\langle d_{nc} \rangle} \exp{(-d_{nc}/\langle d_{nc} \rangle)}$, respectively, with $\langle d_c \rangle  = 1.95$\,mm and $\langle d_{nc} \rangle = 1.46$\,mm. We also plot the probability of the distance travelled by the head in the radial direction $\Delta r_h$ while not in contact with the boundary in Fig.~\ref{fig:scale}(c). It is also described with an exponential function $P(\Delta r_{h}) = \frac{1}{\langle \Delta r_{h} \rangle} \exp{(-\Delta r_{h}/\langle \Delta r_{h} \rangle)}$, with $\langle \Delta r_{h} \rangle =  1.14$\,mm. 
\vspace{0.04in}

We interpret these observed distributions of head travel distances as arising due to the dual strokes used  by the worm to move with amplitudes $A_L$ and $A_T$. The worm performs rapid peristaltic strokes along the length of its body, and somewhat slower transverse undulatory strokes with larger amplitudes~\cite{kudrolli2019burrowing}. Depending on their oscillation phases, the head may slide along as it comes in contact with the boundary, or lose contact with the surface over a distance related to the worm's longitudinal oscillation amplitude $A_L$ and transverse oscillation amplitude $A_T$.
\vspace{0.05in}

Turning to the overall motion of the worm along the boundary, the plot of $\theta_h$ in Fig.~\ref{fig:theta_psi}(d) shows that the worm on average always rotates counter-clockwise in this case, whereas, the worm changes direction once in Fig.~\ref{fig:theta_psi}(e), and more than once in Fig.~\ref{fig:theta_psi}(f). Further, $\psi_b$ plotted in Fig.~\ref{fig:theta_psi}(g-i) can be observed to be negative, positive, or approximately zero, depending on if $\theta_h$ is on average increasing, decreasing, or broadly constant, respectively. Additionally, rapid small scale fluctuations can be observed over short time scales in each of these plots. These rapid fluctuations have to do with the motion of the head which changes direction over a wider range than the body, even while in contact with the boundary as is seen in the plots of $\psi_h$ in Fig.~\ref{fig:theta_psi}(j-l). Thus, while $\psi_h$ can be observed to fluctuate widely as the head can even point away from the surface over the same time instants, $\psi_b$ can be seen to fluctuate far less and appears to be more important in determining the direction of motion. These fluctuations in the head movement have been previously reported in a quasi-2D setup~\cite{patil2022ultrafast}.
\vspace{0.05in}

Comparing the graphs of $\theta_h$ and $\psi_b$ for the same trial, we observe a correlation between the overall slope of $\theta_h$ over a few seconds time interval and the sign of $\psi_b$, i.e. the data appears to indicate that the direction of motion of the worm along the boundary and its relative inclination to it appear to be correlated. To probe this relation between the direction of the worm's motion and its orientation, we calculate $\dot{\theta_h} =\frac{\Delta\theta_h}{\Delta t}$ over a small time interval $\Delta t = 1$\,s. Plotting $\dot{\theta_h}$ as a function of $\psi_b$ in Fig.~\ref{fig:interaction}(a) over 20 trials, where both counterclockwise and clockwise motion are observed, the data can be observed to broadly increase with increasing $\psi_b$. Whereas, obtaining $\theta_h^{\,i}$, the angle between the displacement of the head over the time frames immediately before it contacts the boundary and the normal at the point of contact with the boundary, and plotting $\dot{\theta_h}$ versus $\theta_h^{\,i}$ in Fig.~\ref{fig:interaction}(b), we observe a scatter of points. This confirms that the direction of motion is uncorrelated with the angle that the head subtends with the normal to the boundary. (Only data for one representative trial is plotted for clarity of presentation.)
\vspace{0.05in}

Then, averaging over each of the various trials, we plot the average of $v_b = \dot{\theta_h} R$ versus $\sin({\langle \psi_b \rangle})$ in Fig.~\ref{fig:interaction}(c), noting that the data can be described by a linear fit. This linear fit shows that the more the worm is aligned with the boundary, the faster it moves. Now, if we assume that the tangential speed is unaffected when the worm body collides with the boundary, then, we may expect a relation $ \langle v_b\rangle = v_w \sin({\langle \psi_b \rangle})$, and thus slope can be identified with the unhindered swimming speed of the worm $v_w$. From the fit, we find $v_w = 2.02$\,mm/s, which is  consistent with worm speeds in unbounded environments~\cite{kudrolli2019burrowing}.

\subsection{Time to Reach Passage and Passage Width Dependence}
\begin{figure}
\centering
\hspace*{-0.9cm} 
\includegraphics[width=0.5\textwidth]{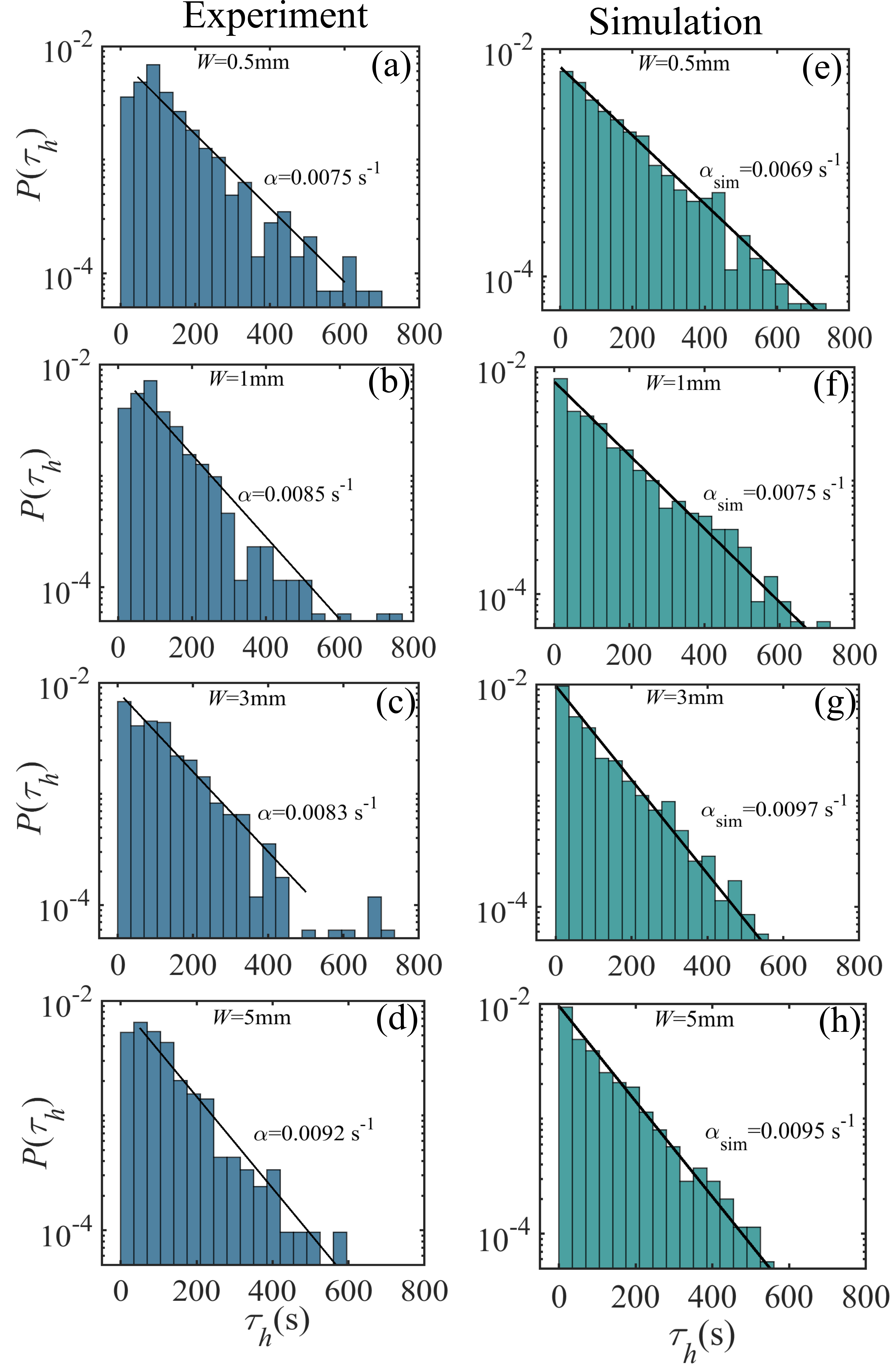}
\caption{\label{fig:tau_h} The measured distribution of time $\tau_h$ taken by the head to enter the exit passage for $W =0.5$\,mm (a), 1\,mm (b), 3\,mm (c), and for 5\,mm (d). The distributions plotted in the right column correspond to the simulations corresponding to the 1D boundary search model, and are observed to show similar exponential distributions.}
\end{figure}

\begin{figure}\centering
\includegraphics[width=0.4\textwidth]{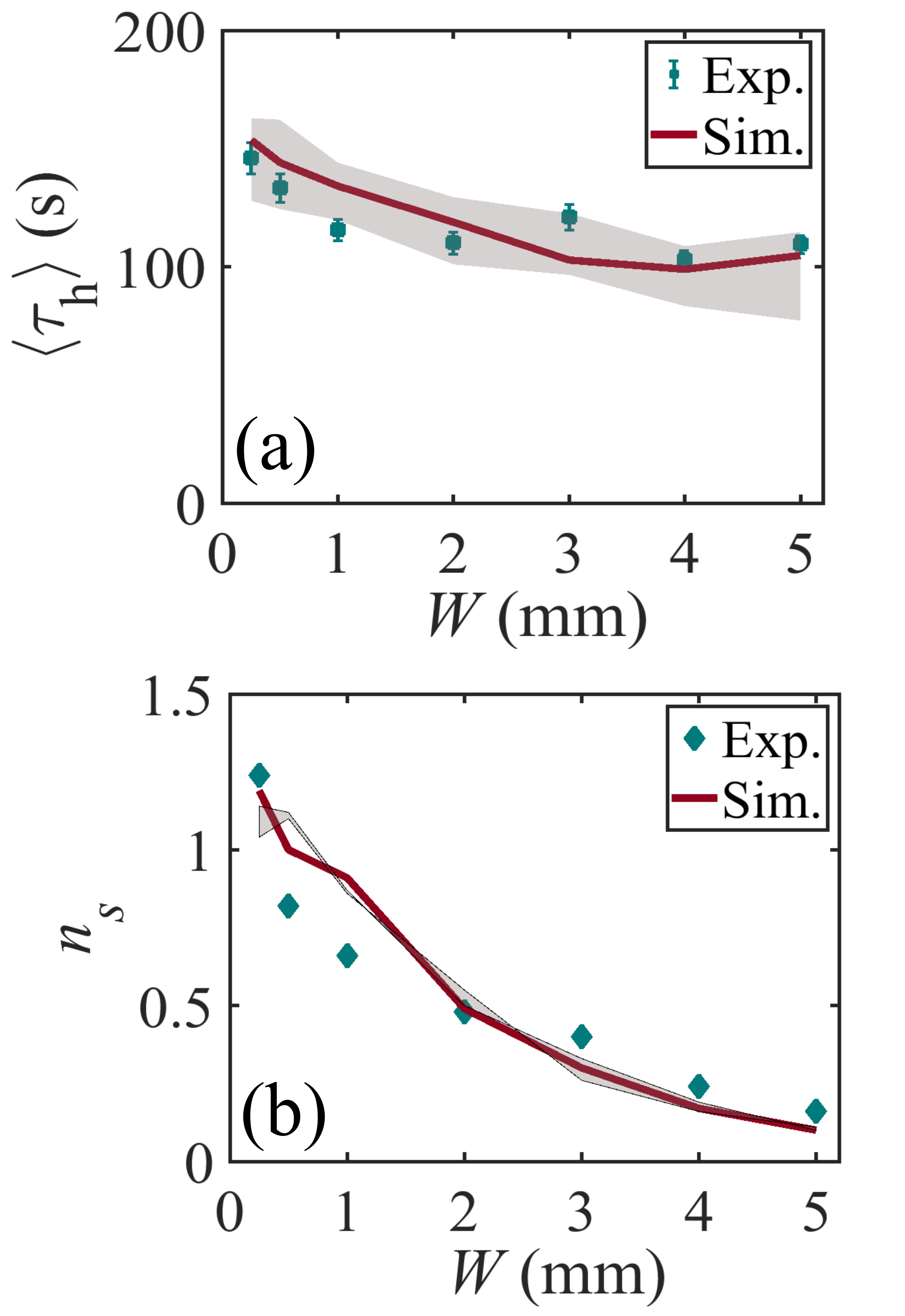}
\caption{\label{fig:sim} (a) The mean time to reach the exit passage as a function of passage width $W$. The overall trends are consistent with measured rate at which the worm’s head enters the exit. Simulation parameters $v_w=2.02$\,mm/s, mean sliding length $1.96$ mm, mean hopping distance $1.47$ mm, are taken from the experiment, and the random direction change is 45\% of the times. (b) The number of times worm skips over the passage $n_s$ decreases with increasing passage width. The shaded regions correspond to varying the rate of random direction change from 35\% to 55\%.}
\end{figure}

Building on the understanding of the interaction of the worm with the boundary, we next examine the time scale over which the worm reaches the exit. We measure the time $\tau_h$ taken by the head to enter the exit passage after entering a chamber, and plot these distributions in Fig.~\ref{fig:tau_h} for various passage width $W$. In each case, $\tau_h$ is broadly distributed and appears to follow an exponential distribution. Plotting the mean time $\langle \tau_h \rangle$ averaged over the 50 trials for each $W$ in Fig.~\ref{fig:sim}(a), we observe that it decreases somewhat, initially, before becoming essentially constant.

If the worm moves without switching directions as it moves around the boundary, then one may expect the time $\tau_h$ over which the worm finds the exit passage to be given by $\tau_h = \frac{R \theta_h^0}{v_w \sin\psi_b}$, where $\theta_h^0$ is the angular position where the head first contacts the boundary while moving clockwise. If the worm moves counter clockwise, $\tau_h = \frac{R (\theta_h^0 - 2\pi)}{v_w \sin\psi_b}$. Assuming $v_w=3$\,mm/s, and an average $\psi_b = 32$\,deg., one can estimate a time $\tau_h \approx 40$\,s  while the worm moves fully around the chamber.
Since the distributions of $\tau_h$ shown are longer time scales, the worm must typically change directions at least a few times as it searches for an exit passage.

\begin{figure}
\centering
{\includegraphics[width=0.3\textwidth]{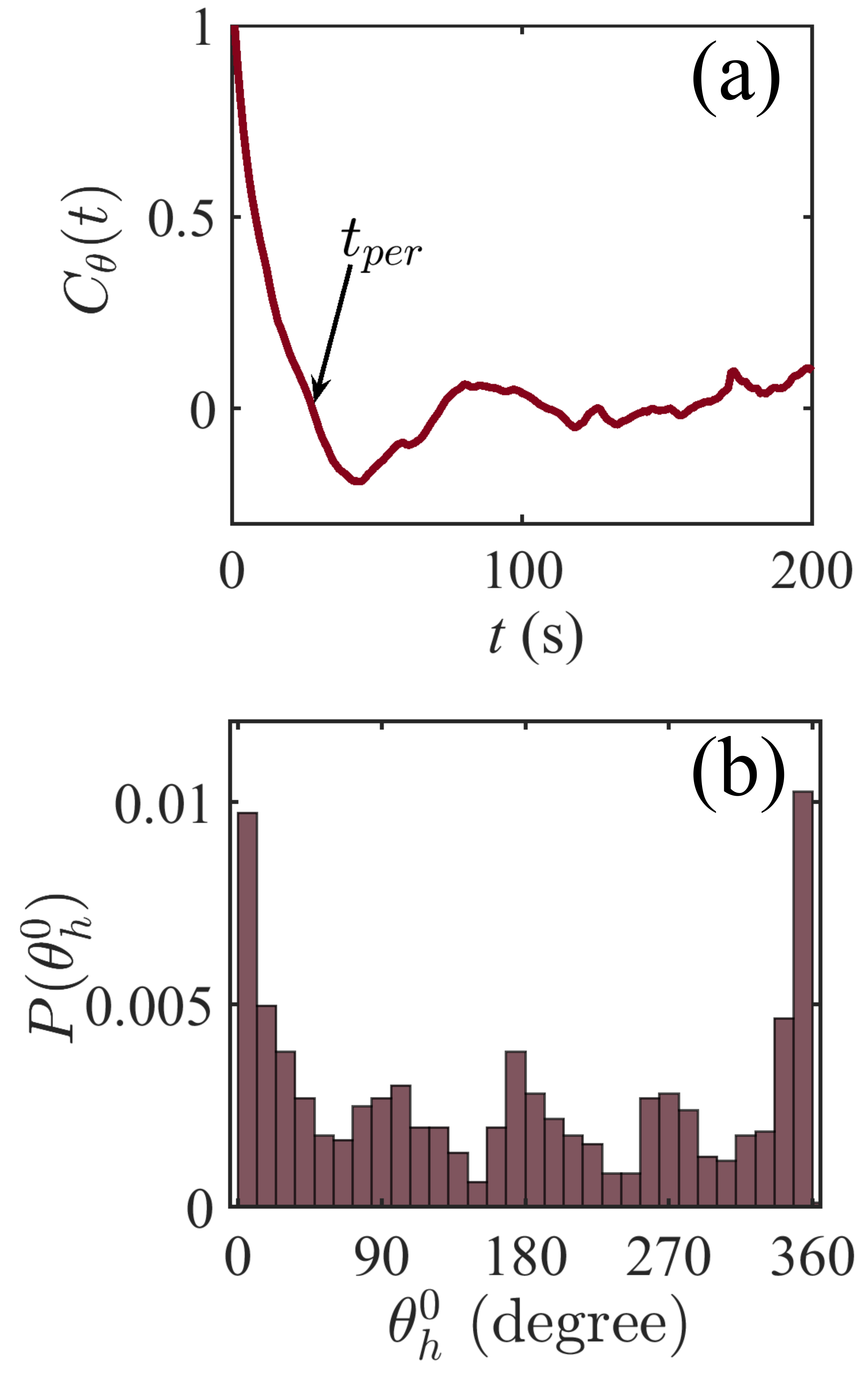}}
\hspace*{0.8cm}    
\caption{\label{fig:vel_corr} (a) The angular correlation function $C_\theta (t)$ of speed ${\dot{\theta_h}}$ versus time $t$. The estimate time scale of persistence motion along boundary is $t_{per}\approx 27$\,s. (b) The distribution of $\theta_h^{0}$ where worm's head initially contacts the boundary after entering the chamber.}
\end{figure}

In order to find the typical time scales over which the worm travels in the clockwise and counterclockwise directions, we calculate the angular correlation function over time $C_\theta (t) = \langle {\rm sgn}(\dot{\theta_h}(t + t_o)) \, {\rm sgn} (\dot{\theta_h} (t_o))\rangle$, where $\langle .. \rangle$ indicates averaging over initial times $t_o$. Figure~\ref{fig:vel_corr}(a) shows a plot of $C_\theta (t)$ where it is observed to decay over a time scale of $t_{per} \approx 27$\,s. Thus, the worm travels in a given direction for a longer time interval compared to its body undulations, and a few changes in directions can be expected statistically while the worm systematically explores the boundary. If the worm's body switches direction as in the situations shown in Fig.~\ref{fig:theta_psi}(h) and Fig.~\ref{fig:theta_psi}(i), the time taken increases with increasing number of switches, giving rise to the longer and wider distributions of $\tau_h$ seen in Fig.~\ref{fig:tau_h}. 

\subsection{Boundary search model to locate exit passage}
In order to understand the time scales over which the worm finds the exit, and the observed variations as a function of its width, we develop a simplified model of its dynamics. Because the worm head plays a dominant role in interacting with the boundary, and it stays near the boundary, we assume that the dynamics can be captured by examining the projected displacement of head along the boundary. Then, we consider the motion along the boundary, when the head is in contact with the boundary, and also when it loses contact with the boundary as being a slip and hop. These slip and hop distances are drawn randomly from an exponential distribution corresponding to their measured distributions shown in Fig.~\ref{fig:scale}(a, b), respectively. The direction is reversed at random, approximately 30\% of the times following the observation that the worm preserves its direction of motion of a time-scale $t_{per}$. Because the experimental geometry is in fact quasi-2D, we take this into account in the simulation by using an effective passage width $w_s = \sqrt{W^2 + h^2}$ in performing these simulations, where $W$ is the width of the channel and $h$ is the thickness of the chamber. Further, we approximate the worm's initial position $\theta_h^0$ on the chamber boundary to have a flat distribution. In the experiments, $\theta_h^0$ is observed to have a systematically higher probability of coming in contact with the boundaries near the passages, i.e. soon after the worms enter the chamber (see Fig.~\ref{fig:vel_corr}(b)).

Figure~\ref{fig:tau_h} shows the numerically calculated distribution of time $\tau_h$, taken to reach the exit, compared with the corresponding measurements from our experiments. We find that they are also exponentially distributed with decay constants increasing slowly with $W$, consistent with the trends observed in the experiments. Comparing the mean time $\langle \tau_h \rangle$ observed in the simulations with those in the experiments in Fig.~\ref{fig:sim}(a), we observe not only good agreement with the overall magnitude of time needed to reach the exit, but also the slightly greater time needed to enter passages that are narrower compared to $A_T$. The increasing trend with $W$, occurs because the worm can skip over the passage while not in contact with boundary as it moves around the chamber. We compare the average times that we observe the worm hops over the passage over a 30 minutes trial, and compare it with the simulations in Fig.~\ref{fig:sim}(b). Excellent quantitative agreement is observed, further validating our model.     

Thus, we find that the worm follows the boundary, doing a persistent random walk in search of the exit, and as a result takes a similar time scale to find the exit, except for widths which are narrower than the distance over which it loses contact with the wall. As we see next, the body undulations which are a natural consequence of its locomotion strokes, can have a further effect on the actual escape time of the worm not only at small exit widths, but wider exits as well. 

\subsection{Effect of Body Undulations on Escape Rate}
\begin{figure}
\centering
\includegraphics[width=0.3\textwidth]{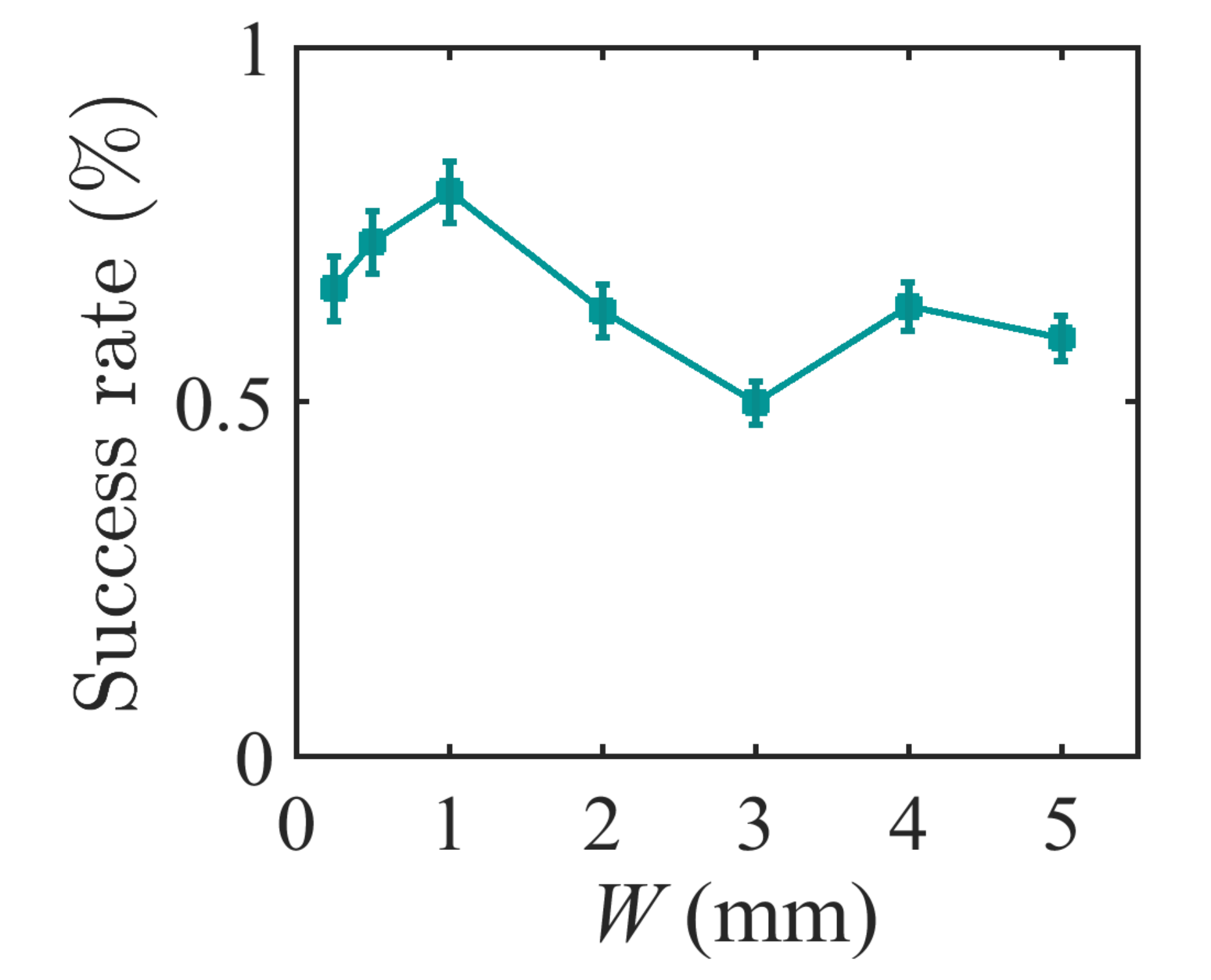}  
\caption{\label{fig:success}  The success rate of worm escape through the passage after its head enters the exit passage.}
\end{figure}

\begin{figure}
\centering
\includegraphics[width=0.5\textwidth]{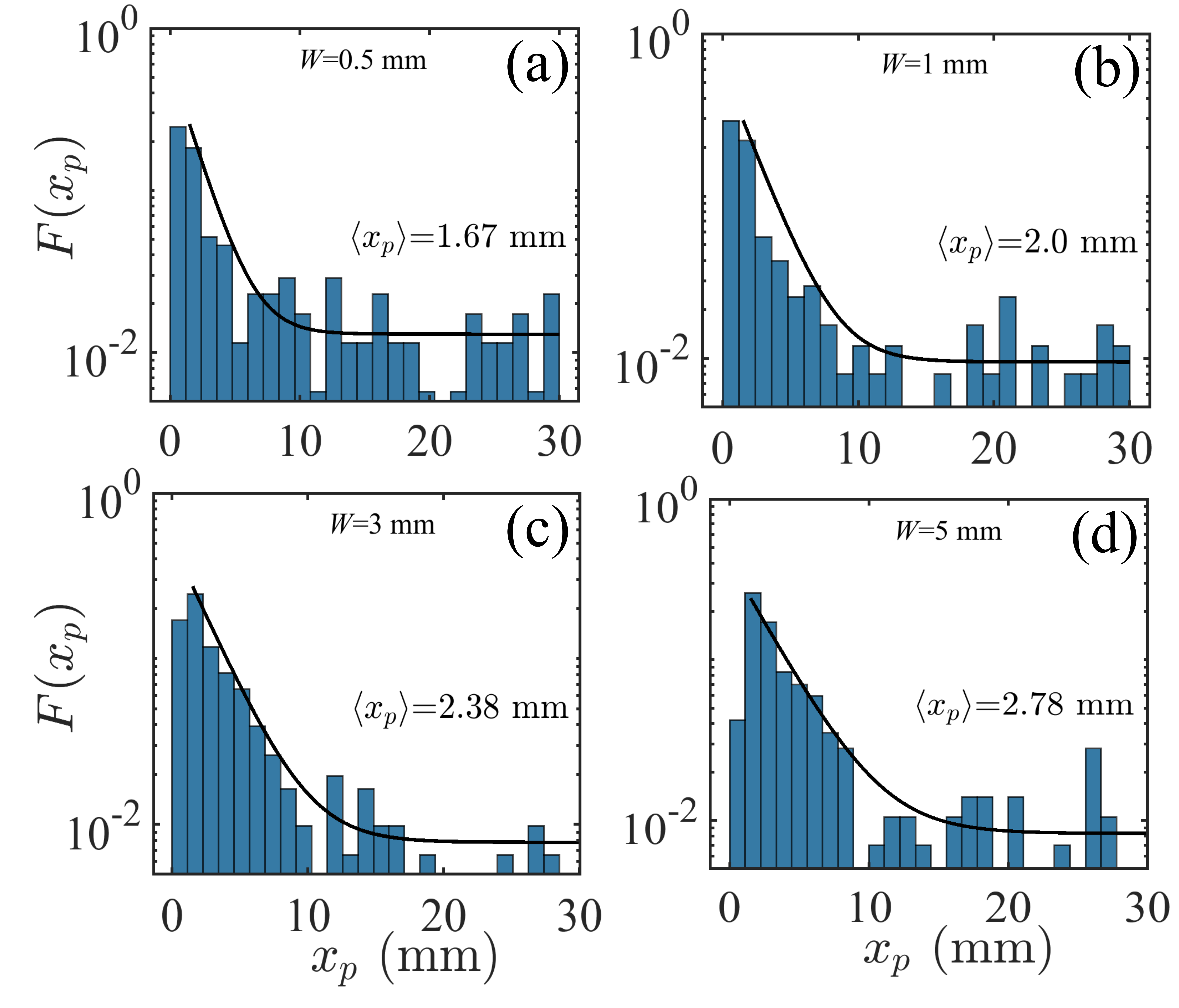}
\caption{\label{fig:pene} The distribution of penetration depth $x_p$, reach by worm before returning back into the chamber for varying widths $W = 0.5$\,mm, $1$ mm, $3$ mm, and $5$ mm. Most of the failed attempts to escape even after entering the passage correspond to the worm's body undulations which leads to the withdrawal of its head involuntarily from the passage.} 
\end{figure}

Figure~\ref{fig:success} shows the success rate of the worm exiting the chamber through the passage once its head has entered into the exit passage, given by the percentage of times that the center of mass of the worm enters the passage, after the head enters the passage. While systematic variation with exit width is difficult to glean given the statistical variation, it is clear that the worm travels through the passage only about 60\% of the times. i.e., the worm fails to recognize that it has entered the passage around 40\% of the times, resulting in failed opportunities to escape, and leading to escape times which are in fact nearly twice as long as compared to $\langle \tau_h \rangle$.

To understand why, we analyze the fraction of times $F(x_p)$ a worm reaches a given penetration distance $x_p$ into the passage along the $x$-axis, when it fails to escape through the exit passage. Figure~\ref{fig:pene} shows the plot of $F(x_p)$ for various passage widths $W$. In each case, we observe that the worm's head typically enters the passage for a very short distance before returning back into the chamber. In particular, we can describe $F(x_p)$ as a sum of an exponential decaying function and a constant ($F(x_p) = \alpha \exp{(-x_{p}/\langle x_{p} \rangle}$), which corresponds to a small but constant probability for the worm turning back in a passage, rather than proceeding forward. By fitting, we find the exponential decay constant $\beta=1/\langle x_{p} \rangle = 0.60$\,mm$^{-1}$, 0.50\,mm$^{-1}$, 0.42\,mm$^{-1}$, and 0.36\,mm$^{-1}$, for $W = 0.5$\,mm, 1\,mm, 3\,mm, and 5\,mm, respectively. Thus, the penetration into the passage increases somewhat with $W$, but still appears to be of order of the stroke amplitudes. 

Consequently, we attribute the missed opportunities to escape through the exit passage to the undulations of the body used in locomotion, whereby the worm's head enters the exit passage and then subsequently withdraws back into the chamber. To understand this behavior further, we examine the dynamics of the worm in each case where the worm failed to escape through the passage for $W = 0.5$\,mm, 1\,mm, and 3\,mm. While a total of 79, 98, and 222 failed entries, respectively, were recorded, the percentage of times the worm had no contact with the wall when failing to fully exit increased from about 2\% to about 20\% with increasing width. Whereas, the worm came in contact with the boundaries during the elongation phase of its peristaltic stroke, 93\%, 87\%, and 53\% of the times, respectively, before its head withdraws back into the chamber, leading to failed escapes. Thus, 99\%  of the reason why the worm fails to fully enter the exit passage in the case of the narrowest channel, and about 73\% in the case of the wider ($W=3$\,mm) channel  can be directly attributed to its natural undulations. In the remaining cases, it appears that the worm comes into contact with the boundaries, but fails to recognize that it is in the exit channel. This appears to indicate that more than one contact is required within a stroke cycle before the worm recognizes that it is inside a passageway.

\section{Conclusions}
In summary, by constructing transparent quasi-2D chambers that enable us to track the motion of the entire shape of the worm over time, we have shown that {\it Lumbriculus variegatus} use a boundary following strategy to navigate enclosed spaces. Although the worms have sensory nerves all over their body, we find that the worm primarily uses prostomial nerves located in its head to actively probe the boundary. We then show with statistical analysis that the worm's  direction of motion is determined by the orientation of its entire body, rather than its head. While the time to find the exit passage would increase linearly with passage width, if the worm were to explore the chamber ballistically, we show that the worm's head enters the passage roughly over the same time scale even as the width of the exit passage is varied over an order of magnitude. A slightly greater time is needed to find the passages at the narrowest widths because the swimming strokes lead the worms to lose contact with the boundary and skip over the passage opening. This observation further provides evidence that the worm locates the boundaries and exit passage due to steric interactions during contact, and that long range interactions mediated by hydrodynamic flows are not important to its search strategy. The thigmotaxis or hiding behavior of these worms does not appear to have an effect on finding the exit passage, under the conditions studied.

We further show that we can capture the time scales over which the worm reaches the exit using a one-dimensional persistent random walk model of its trajectory along the boundary. This model captures the mean time taken and their distributions as the worm does not always search the boundary in one direction, but also randomly reverses directions to retrace its path. This enables the worm to find a passage which it may have missed without fully circling the enclosure because it is not always in contact with the boundary while probing for a path forward. This can be noted to be especially important when the passages have small widths compared to the mean longitudinal peristaltic and transverse undulation amplitudes. The strokes can further lead the worm to miss the exit passage even after its head enters the passage, as they can cause it to withdraw its head back into the chamber. While our persistent random walk model is able to capture the escape time dynamics of the worm’s head, a more elaborate two-dimensional active model which takes into account the undulatory and peristaltic strokes of the worm is needed to further capture the detailed dynamics of the entire body.

Our study shows an efficient strategy employed by a flexible limbless intruder to actively search on a surface for openings using only sensory nerves located at one end of its body. While boundary following has been shown in microscopic motile organisms including sperm, bacteria, algae and nematodes~\cite{Park2003,park2008enhanced,denissenko2012human,yuan2015hydrodynamic,kantsler2013ciliary}, our study provides a thorough examination of the body shape and strokes and demonstrate their importance on the navigation strategies employed by organisms using the sense of touch. The one-dimensional boundary search model introduced here is quite general and may be applicable to other animal systems where boundary following by direct contact is important~\cite{creed1990interpreting}.

\begin{acknowledgments}
We acknowledge the support of the U.S. NSF grant CBET-1805398, and thank Professor Alex Petroff for allowing access and guidance to the Laser-cutter facility in his laboratory.

\end{acknowledgments}
\bibliographystyle{rsc}
\bibliography{worms}
\end{document}